\documentclass[aps,twocolumn,showpacs,pre]{revtex4-1}
\usepackage{graphicx,graphics}
\usepackage{amsmath,amssymb,amsfonts}
\usepackage{multirow}
\usepackage{hyperref}
\usepackage{graphicx}
\usepackage{xcolor}

\begin{document}
\title{Optimal exploration of random walks with local bias on networks}
\author{Christopher Sebastian Hidalgo Calva$^1$}
\author{Alejandro P. Riascos$^2$}
\affiliation{${}^1$Instituto de Investigaciones en Matem\'aticas Aplicadas y Sistemas, Universidad Nacional Aut\'onoma de M\'exico, Mexico City, 04510, Mexico}
\affiliation{${}^2$ Instituto de F\'isica, Universidad Nacional Aut\'onoma de M\'exico, 
		Apartado Postal 20-364, 01000 Ciudad de M\'exico, M\'exico}

\begin{abstract}
We propose local-biased random walks on general networks where a Markovian walker is defined by different types of biases in each node to establish transitions to its neighbors depending on their degrees. For this ergodic dynamics, we explore the capacity of the random walker to visit all the nodes characterized by a global mean first passage time. This quantity is calculated using eigenvalues and eigenvectors of the transition matrix that defines the dynamics. In the first part, we illustrate how our framework leads to optimal exploration for small-size graphs through the analysis of all the possible bias configurations. In the second part, we study the most favorable configurations in each node by using simulated annealing. This heuristic algorithm allows obtaining approximate solutions of the optimal bias in different types of networks. The results show how the local bias can optimize the exploration of the network in comparison with the unbiased random walk. The methods implemented in this research are general and open the doors to a broad spectrum of tools applicable to different random walk strategies and dynamical processes on networks.
\end{abstract}

\maketitle
\section{Introduction}
\begin{figure}[!b]
\begin{center}
\includegraphics*[width=0.5\textwidth]{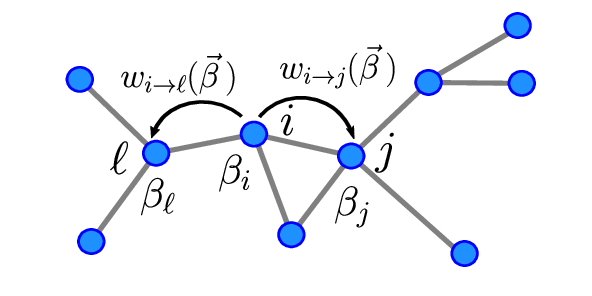}
\end{center}
\vspace{-5mm}
\caption{\label{Fig_1} A random walker with local bias illustrated as different transition probabilities between the nodes in a network. The possible movements from a node $i$ to its neighbors are given by transition probabilities defined by $\vec{\beta}=(\beta_1,\beta_2,\ldots,\beta_N)$, where $N$ is the size of the network and $\beta_s$ a parameter associated to each node $s$. The walker can modify each element of $\vec{\beta}$  trying to discover a strategy that optimizes its capacity to reach all the nodes of the network.}
\end{figure}

Markovian random walks on networks are found in different contexts and have been studied intensively over the last decades, from a purely theoretical perspective \cite{Hughes,MasudaPhysRep2017,FractionalBook2019,ReviewJCN_2021} to diverse applications such as database exploration \cite{LeskovecBook2014,BlanchardBook2011}, web page ranking \cite{GooglePR1998,ShepelyanskyRevModPhys2015}, epidemic spreading \cite{Bestehorn2021}, encounter networks \cite{RiascosMateosPlosOne2017}, human mobility \cite{RiascosMateosSciRep2020},  among many others. These dynamics have been implemented in diverse fields as processes that are able to efficiently reach hidden targets or to simply explore a particular region of space \cite{ReviewJCN_2021}.
\\[2mm]
In particular, degree-biased random walks describe a dynamics where a walker in a particular network node uses the degree information of neighbors to choose randomly a new node to visit \cite{YangPRE2005,WangPRE2006,FronczakPRE2009}; in this case, a tunable parameter $\beta$ controls the bias in the complete network.  This process is of interest in theory and applications to understand how $\beta$ affects the exploration of a network in different topologies \cite{YangPRE2005,WangPRE2006,FronczakPRE2009,ReviewJCN_2021}. The spectrum of applications includes traffic dynamics and routing protocols \cite{WangPRE2006}, the study of simultaneous random walkers \cite{WengPRE2017,WengChaos2018,WengPRE2018universal,RiascosSandersPRE2021}, extreme events \cite{KishorePRE2012}, epidemic spreading \cite{PUPhysA2015}, just to mention a few examples. Recently, degree-biased random walks have been generalized to include multiple biases \cite{Wang_Chaos2021}, in potential-driven random walks  \cite{BenigniPRE2021}, to study the influence of damage and aging in complex systems  \cite{Aging_PhysRevE2019,Eraso_Hernandez_2021}.
\\[2mm]
In different studies, degree-biased random walks are defined with a global parameter $\beta$; in this manner, its value is the same for all the nodes in the network. However, this parameter could be local, i.e., with values in each node. This type of dynamics is illustrated in Fig. \ref{Fig_1}. In this example, the transition probability from a node $i$ to one of its neighbors depends on a parameter $\beta_i$. Hence, the transition probability matrix that defines the random walker is determined by a vector $\vec{\beta}=(\beta_1,\beta_2,\ldots,\beta_N)$, being $N$ the size of the network. In connected undirected networks, finite values of $\beta_i$ define a dynamics capable of reaching any node from any initial condition. 
\\[2mm]
In this paper, we explore degree-biased random walk dynamics with local bias defined by a vector $\vec{\beta}$. In Sec. \ref{RWdefinitions}, we present a summary of the formalism to obtain mean-first passage times and its weighted average over all the nodes to describe the transport of ergodic random walkers on networks.
In Sec. \ref{LocalBiasGeneral}, we introduce the formal definitions of random walks with a local bias and the motivation for this stochastic process. We analyze general properties and the optimal dynamics in graphs with small sizes ($N=4,5$) through the exploration of all the possible bias configurations. In Sec. \ref{Optimal_Trans}, we analyze networks with different sizes and topologies using simulated annealing. This heuristic method allows us to efficiently identify particular values of the local bias that are most favorable (or close to an optimal) to reach all the nodes of the network. We explore this algorithm in different types of networks. The framework introduced in this research is general and can be extended to other random walk strategies with local parameters to define the dynamics. This approach may have applications in routing processes, to address common problems in traffic planning in public transportation, among other cases where it is necessary to tune parameters in the parts of a complex system.
\section{Random walks on networks}
\label{RWdefinitions}
\subsection{Random walks and mean first passage times}
\label{App_DeductionTij}
In this section, we discuss the mean first passage times (MFPTs) and a global time to characterize random walks on networks. We consider the dynamics on connected  undirected networks with $N$ nodes $i=1,\ldots ,N$ described by an adjacency matrix $\mathbf{A}$ with elements $A_{ij}=A_{ji}=1$ if there is an edge between the nodes $i$ and $j$ and $A_{ij}=0$ otherwise; in particular, $A_{ii}=0$ to avoid edges connecting a node with itself. The degree of a node $i$ is given by $k_i=\sum_{\ell=1}^N A_{i\ell}$, this is the number of neighbors of $i$.
\\[2mm]
The occupation probability to find a Markovian random walker in a node $j$ at time $t$, starting 
from $i$ at $t=0$, is given by $P_{ij}(t)$ and obeys the master equation \cite{Hughes,ReviewJCN_2021,NohRieger2004}
\begin{equation}\label{master}
P_{ij}(t+1) = \sum_{\ell=1}^N  P_{i\ell} (t) w_{\ell\rightarrow j} \ ,
\end{equation}
where we consider discrete times $t=0,1,\ldots$. In Eq. (\ref{master}), $w_{\ell\to j}$ is the transition probability to randomly hop in one step from node $\ell$ to $j$. These probabilities define a $N\times N$ matrix $\mathbf{W}$ with elements $\mathbf{W}_{ij}=w_{i\to j}$. All the information of the dynamics is defined by $\mathbf{W}$. In the following, we study the dynamics with transition matrices $\mathbf{W}$ for which the random walker can reach in finite time any node starting from any initial condition; in this manner, $\mathbf{W}$ defines an ergodic process.
\\[2mm]
Using  the probability $P_{ij}(t)$ and the stationary distribution $P_j^\infty\equiv \lim_{T\to \infty} \frac{1}{T}\sum_{t^\prime=0}^T P_{ij}(t^\prime)$, the moment $R^{(0)}_{ij}$ is defined as \cite{NohRieger2004}
\begin{equation}
	R^{(0)}_{ij}\equiv \sum_{t=0}^{\infty}\{P_{ij}(t)-P_j^\infty\}.
\end{equation}
In terms of these quantities, the analytical form of the MFPT $\left\langle  T_{ij}\right\rangle$ that gives the average number of steps to start in $i$ and reach for the first time the node $j$ is \cite{NohRieger2004}
\begin{equation}\label{Tij}
	\langle T_{ij} \rangle =\frac{1}{P_j^\infty}\left[R^{(0)}_{jj}-R^{(0)}_{ij}+\delta_{ij}\right] .
\end{equation}
In order to calculate $\langle T_{ij} \rangle$ it is necessary to find $P_{ij}(t)$. In the following, we use Dirac's notation, then
\begin{equation}\label{ProbVector}
P_{ij}(t)=\left\langle i\right|\mathbf{W}^t \left|j\right\rangle,
\end{equation}
where $\{\left|m\right\rangle \}_{m=1}^N$ represents the canonical base of $\mathbb{R}^N$. 
\\[2mm]
For ergodic random walks on connected undirected networks, the matrix $\mathbf{W}$ can be diagonalized. For right eigenvectors of $\mathbf{W}$ we have $\mathbf{W}\left|\phi_i\right\rangle=\lambda_i\left|\phi_i\right\rangle $ 
for $i=1,\ldots,N$, where the eigenvalues $\lambda_i$ can take complex values  and are sorted in the form $\lambda_1=1$ and $1\geq|\lambda_l|\geq 0$ for $l=2,3,\ldots, N$. With this information, we define a matrix $\mathbf{Z}$ with elements $Z_{ij}=\left\langle i|\phi_j\right\rangle$ \cite{ReviewJCN_2021}. The matrix $\mathbf{Z}$ is invertible, and a new set of vectors $\left\langle \bar{\phi}_i\right|$ is obtained by means of $(\mathbf{Z}^{-1})_{ij}=\left\langle \bar{\phi}_i |j\right\rangle $, then
\begin{equation*}
\delta_{ij}=(\mathbf{Z}^{-1}\mathbf{Z})_{ij}=\sum_{l=1}^N \left\langle\bar{\phi}_i|l\right\rangle \left\langle l|\phi_j\right\rangle=\langle\bar{\phi}_i|\phi_j\rangle \, 
\end{equation*}
and $\mathbb{I}=\mathbf{Z}\mathbf{Z}^{-1}=\sum_{l=1}^N \left|\phi_l\right\rangle \left\langle \bar{\phi}_l \right| $, where $\mathbb{I}$ is the $N\times N$ identity matrix. 
\\[2mm]
By using the spectral form of $\mathbf{W}$,  Eq. (\ref{ProbVector}) takes the form
\begin{equation}\label{PtSpect}
P_{ij}(t)=\sum_{l=1}^N\lambda_{l}^t\left\langle i|\phi_l\right\rangle \left\langle \bar{\phi}_l|j\right\rangle  \, .
\end{equation}
From Eq. (\ref{PtSpect}), the stationary probability distribution is $P_j^{\infty}=\left\langle i|\phi_1\right\rangle \left\langle \bar{\phi}_1|j\right\rangle$, where the result $\left\langle i|\phi_1\right\rangle=\rm{constant}$ makes $P_j^{\infty}$ independent of the initial condition. Now, as a consequence of the definition of $R_{ij}^{(0)}$ \cite{ReviewJCN_2021}
\begin{equation}\label{RijSpect}
R_{ij}^{(0)}=\sum_{l=2}^N\frac{1}{1-\lambda_l}\left\langle i|\phi_l\right\rangle \left\langle\bar{\phi}_l|j\right\rangle \, .
\end{equation}
Therefore, for $i \neq j$ in  Eq. (\ref{Tij}), the MFPT $\left\langle T_{ij}\right\rangle$ is \cite{ReviewJCN_2021}
\begin{equation}\label{TijSpect}
\left\langle T_{ij}\right\rangle
=\sum_{l=2}^N\frac{1}{1-\lambda_l}\frac{\left\langle j|\phi_l\right\rangle \left\langle\bar{\phi}_l|j\right\rangle-\left\langle i|\phi_l\right\rangle \left\langle\bar{\phi}_l|j\right\rangle}{\left\langle j|\phi_1\right\rangle \left\langle\bar{\phi}_1|j\right\rangle}\, ,
\end{equation}
whereas $\left\langle T_{ii}\right\rangle=(\left\langle i|\phi_1\right\rangle \left\langle\bar{\phi}_1|i\right\rangle)^{-1}$ is the mean first return time to the node $i$. In addition, we have the time $\tau_j$ independent of the initial condition \cite{RiascosMateos2012,ReviewJCN_2021}
\begin{equation}\label{TauiSpect}
\tau_j\equiv\sum_{l=2}^N\frac{1}{1-\lambda_l}\frac{\left\langle j|\phi_l\right\rangle \left\langle\bar{\phi}_l|j\right\rangle}{\left\langle j|\phi_1\right\rangle \left\langle\bar{\phi}_1|j\right\rangle}\, .
\end{equation}
Using this definition, we have 
\begin{equation}
\sum_{i=1}^N P_{i}^{\infty}\left\langle T_{ij}\right\rangle=1 +\tau_j,
\end{equation}
where we use the relation $\sum_{i=1}^N P_i^\infty \left\langle i|\phi_l\right\rangle=0$ valid for $l=2,3,\ldots N$, this is a consequence of the relation $\left\langle\bar{\phi}_1|\phi_\ell\right\rangle=\delta_{1\ell}$. In this manner $\tau_j=\sum_{i\neq  j}P_{i}^{\infty}\left\langle T_{ij}\right\rangle$  
characterizes the capacity of the dynamics to reach the node $j$ from any initial condition $i\neq j$ (see Ref. \cite{NohRieger2004} for additional details). Hence, this value allows defining a global time $\mathcal{T}$ to quantify the exploration of the whole network \cite{RiascosMateos2012,Eraso_Hernandez_2021}
\begin{equation}\label{TauGlobal}
\mathcal{T}\equiv\frac{1}{N}\sum_{j=1}^N\tau_j=\frac{1}{N}\sum_{j=1}^N\sum_{i\neq j}P_{i}^{\infty}\left\langle T_{ij}\right\rangle .
\end{equation}
In this relation, we see that $\mathcal{T}$  is a global time that gives the weighted average of the number of steps to reach any node of the network. In the following, we use the global time  $\mathcal{T}$  to quantify the capacity of the random walker to explore the network using the eigenvalues and eigenvectors of the transition matrix \cite{ReviewJCN_2021,FractionalBook2019}. Other alternatives for a global characterization of the dynamics may include mixing times of Markov chains \cite{BlanchardBook2011} or the cover time defined as the time a random walker requires to visit every node in the network at least once \cite{Brockmann2017}.
\subsection{Degree-biased random walks}
\label{DBRWsection}
In this section, we discuss degree-biased random walks. For this case, the random walker hops with transition probabilities $w_{i \to j}$ depending on the degrees of the neighbors of the node $i$. Degree-biased random walks are defined by \cite{FronczakPRE2009} 
\begin{equation}\label{wijBRWglobal}
	w_{i\rightarrow j}=\frac{A_{ij} k_j^{\beta}}{\sum_{l=1}^N A_{il} k_l^{\beta}},
\end{equation}
where $\beta$ is a real parameter. In Eq.~(\ref{wijBRWglobal}), $\beta>0$ describes the bias to hop to neighbor nodes with a higher degree, whereas for $\beta<0$ this behavior is inverted and, the walker tends to hop to nodes less connected. When $\beta=0$, the normal random walk with $w_{i\to j}=A_{ij}/k_i$ is recovered.
\\[2mm]
The random walk in Eq. (\ref{wijBRWglobal}) is also defined in terms of a symmetric matrix of weights $\mathbf{\Omega}$ with elements $\Omega_{ij}=A_{ij}(k_i k_j)^\beta$ with transition probabilities \cite{ReviewJCN_2021}
\begin{equation*}
    w_{i\rightarrow j}=\frac{\Omega_{ij}}{\sum_{l=1}^N \Omega_{il}}=\frac{\Omega_{ij}}{S_i}.
\end{equation*}
Here $S_i=\sum_{l=1}^N \Omega_{il}$ and represents the total weight of the node $i$ (see Ref. \cite{ReviewJCN_2021} for a review of different random walks defined using a symmetric matrix of weights). In terms of this formalism, in connected undirected networks, degree  biased random walks are ergodic for $\beta$ finite and the stationary distribution is given by \cite{ReviewJCN_2021}
\begin{equation}\label{StatPBRW}
	P_i^{\infty}=\frac{S_i}{\sum_{l=1}^N S_l}=\frac{\sum_{l=1}^N (k_i k_l)^\beta A_{il}}{\sum_{l,m=1}^N (k_l k_m)^\beta A_{lm}} \, .
\end{equation}
Degree  biased random walks have been studied extensively in the literature in different contexts as varied as routing processes \cite{WangPRE2006}, chemical reactions \cite{KwonPRE2010}, extreme events \cite{KishorePRE2012,LingEPJB2013}, multiple random walks on networks \cite{WengPRE2017,RiascosSandersPRE2021}, among others \cite{FronczakPRE2009,LambiottePRE2011,Battiston2016}. Additionally, mean field approximations have been explored for diverse cases \cite{FronczakPRE2009,KwonPRE2010,ZhangJSMTE2011}.
\\[2mm]
\begin{figure}[!t] 
	\begin{center}
		\includegraphics*[width=0.47\textwidth]{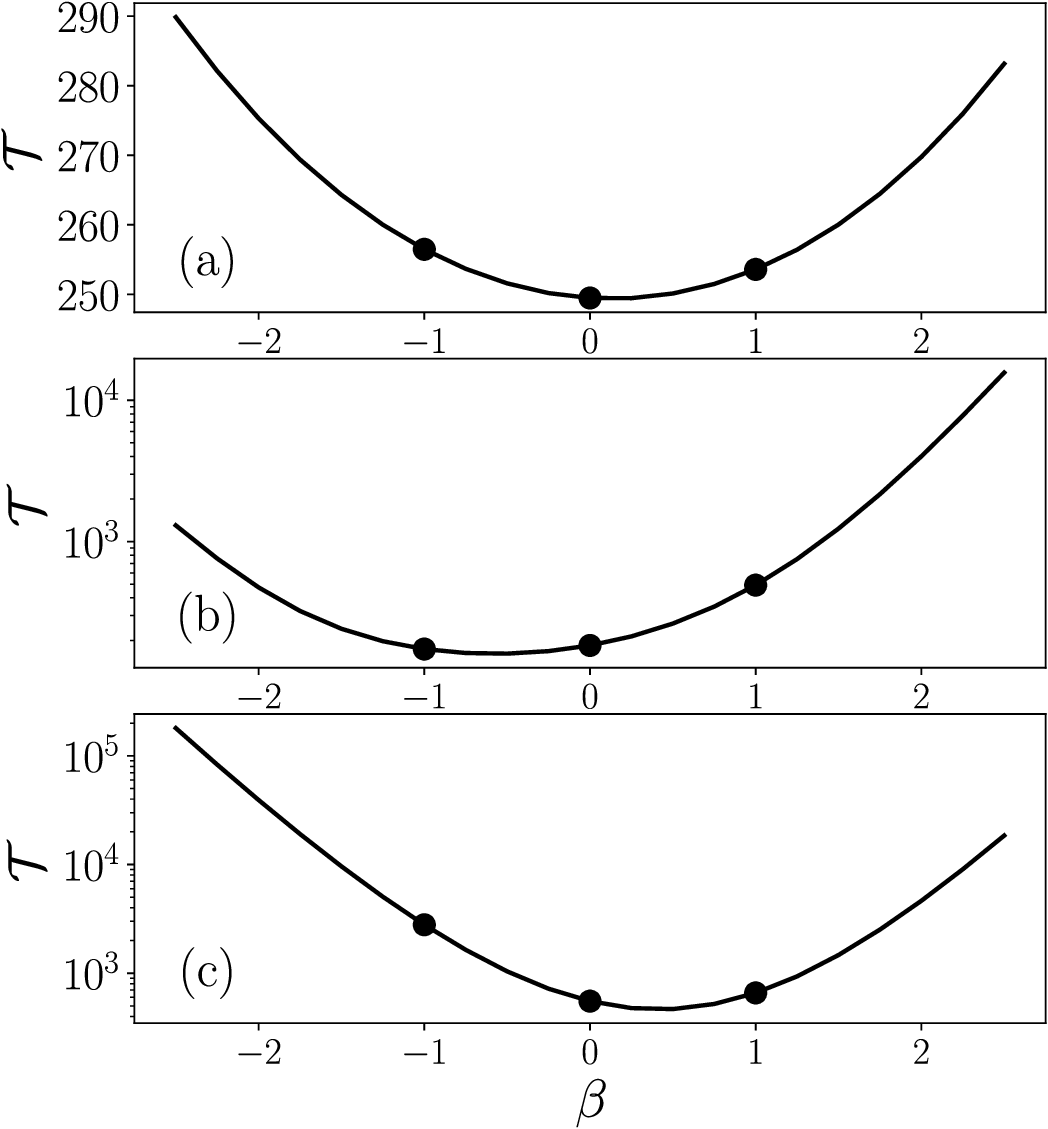}
	\end{center}
	\vspace{-5mm}
	\caption{\label{Fig_2} Biased random walks on networks with $N=100$ nodes: (a) Erd\H{o}s-R\'enyi, (b)  Watts-Strogatz, (c) Barab\'asi-Albert network (details in the main text). We show the average global time $\mathcal{T}$ obtained numerically using Eq.~(\ref{TauGlobal}) for  degree-biased random walks defined by the transition probabilities in Eq.~(\ref{wijBRWglobal}) with $\beta=-2.5,-2.25,\ldots,2.25,2.5$ (continuous lines are used as a guide). The values with dots represent the results for the discrete values $\beta=-1,0,1$. } 
\end{figure}
In Fig. \ref{Fig_2} we show the global time $\mathcal{T}$ in Eq.~(\ref{TauGlobal}) for degree-biased random walkers defined by Eq. (\ref{wijBRWglobal}) as a function of $\beta$. We examine the results for the dynamics on three random structures with $N=100$: an Erd\H{o}s-R\'enyi network \cite{ErdosRenyi1959} with probability $p=\ln(N)/N$ to define the lines, a Watts-Strogatz network with rewiring probability $p=0.05$ \cite{WattsStrogatz1998} and  a scale-free Barab\'asi-Albert network, generated with the preferential attachment rule where each newly introduced node connects to $m$ previous nodes, we choose $m=1$ \cite{BarabasiAlbert1999}. The results are obtained numerically for $\beta\in[-2.5,2.5]$, our findings illustrate how the parameter $\beta$ modifies the global exploration of the network in different topologies.  In particular, in Fig. \ref{Fig_2}(a) we see that an optimal exploration is obtained for $\beta=0$, whereas in Fig. \ref{Fig_2}(b) the random walker may benefit using the bias $\beta=-1$ that gives preference to the pass to less connected nodes. On the contrary, for the complex network in Fig. \ref{Fig_2}(c), the best choice is to use a small bias towards highly connected nodes, for example, implementing a random walk with $\beta\in (0,1)$. 
\\[2mm]
In addition, regarding the values of $\mathcal{T}$ to characterize the dynamics on heterogeneous networks, for $|\beta|\gg 1$ the dynamics with bias requires a huge number of steps to reach any node since the random walker gets trapped with revisits to the highly connected nodes (for $\beta\gg1$) or in nodes with reduced connectivity (for  $\beta\ll -1$)  yielding an optimum for finite values $\beta$.
\section{Random walks with local bias}
\label{LocalBiasGeneral}
\subsection{General definition}
In the case presented in Eq. (\ref{wijBRWglobal}), the parameter $\beta$ defines the global bias of the random walker. However, another alternative is to choose this parameter at each node; in this manner, by using a vector $\vec{\beta}=(\beta_1,\beta_2,\ldots,\beta_N)$ with the {\it local bias} $\beta_i\in \mathbb{R}$ at node $i$, we define
\begin{equation}\label{wijBRWlocal}
w_{i\rightarrow j}(\vec{\beta}\,)=\frac{A_{ij} k_j^{\beta_i}}{\sum_{l=1}^N A_{il} k_l^{\beta_i}}.
\end{equation}
\begin{figure*}[t!]
	\begin{center}
		\includegraphics[width=1.0\linewidth]{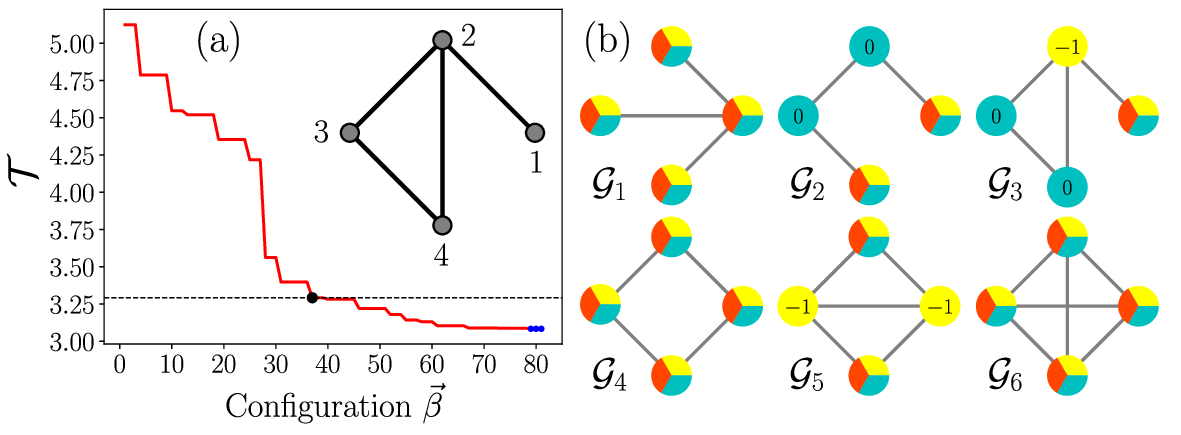}
		\caption{ \label{Fig_3}  Random walks with local bias on graphs with $N=4$. (a) Global times $\mathcal{T}$ for the 81 configurations of $\vec{\beta}$ with $\beta_i\in \{-1,0,1\}$ for the particular graph presented as inset. The results are sorted in decreasing order and the dashed line represents the value $\mathcal{T}_{\mathrm{unbiased}}$ obtained for $\vec{\beta}=(0,0,0,0)$. (b) Non-isomorphic connected networks with $N=4$. In this case, the nodes represent the multiple values that $\beta_i$ can take for the optimal configuration (with minimal  $\mathcal{T}$), nodes shown as a pie chart represent $\beta_i\in \{-1,0,1\}$, in other nodes, the bias $\beta_i$ is unique with its value included in the respective node.}
	\end{center}
\end{figure*}
Then, the vector $\vec{\beta}$ and the network structure defines the random walk with local bias, a strategy that allows to control at each node the transitions from a node to more or less connected neighbors.  Note that, when $\vec{\beta}=(\beta, \beta,\ldots, \beta)$, we recover the random walk with global bias discussed in Sec. \ref{DBRWsection}.
\\[2mm]
In particular, for finite values of $\beta_i$, the random walker with transition matrix $\mathbf{W}(\vec{\beta})$ with elements $w_{i\rightarrow j}(\vec{\beta}\,)$ is ergodic in connected undirected networks, i.e. maintains the capacity to reach any node from any initial condition. In the general case, $\mathbf{W}(\vec{\beta})$ is not defined in terms of a symmetric matrix of weights and we cannot deduce explicitly the stationary distribution. However, the results in  Eqs. (\ref{TijSpect})-(\ref{TauGlobal}) are still valid for the analysis of the dynamics with local bias by using the respective eigenvalues and eigenvectors of $\mathbf{W}(\vec{\beta})$. In particular, the eigenvalues $\lambda_i$ can take complex values, with $\lambda_1=1$ unique and $0\leq|\lambda_l|\leq 1$ for $l=2,3,\ldots, N$, in agreement with the Perron-Frobenius theorem (see Refs. \cite{DirectedFractional_PRE2020,Eraso_Hernandez_2021} for two examples of ergodic dynamics with bias and complex eigenvalues).
\\[2mm]
Once defined the dynamics with local bias, the question arises as to whether it is possible to find configurations of the $\beta_i$ or the vector $\vec{\beta}$ to optimize the capacity of the random walker to reach any node of the network. If these optimal strategies exist, are they unique? How do we find the best values? How are the optimal $\vec{\beta}$ related to the network topology?
\\[2mm]
In the following, we explore those questions. For simplicity, we limit the analysis to a case where the values $\beta_i\in \{-1,0,1\}$ representing three types of bias. However, the approach implemented is general and can be applied to other sets of parameters. The capacity to explore the network is evaluated with the global time $\mathcal{T}(\vec{\beta})$ defined in Eq. (\ref{TauGlobal}) and obtained numerically with the eigenvalues and eigenvectors of $\mathbf{W}(\vec{\beta})$. Our main goal is to determine if there exists a reduction of $\mathcal{T}(\vec{\beta})$ in comparison with the dynamics without bias recovered when $\vec{\beta}=(0,0,\ldots,0)$.
\subsection{Dynamics on small graphs}
\label{Sec_Small}
\begin{figure*}[t!]
	\begin{center}
		\includegraphics[width=1\linewidth]{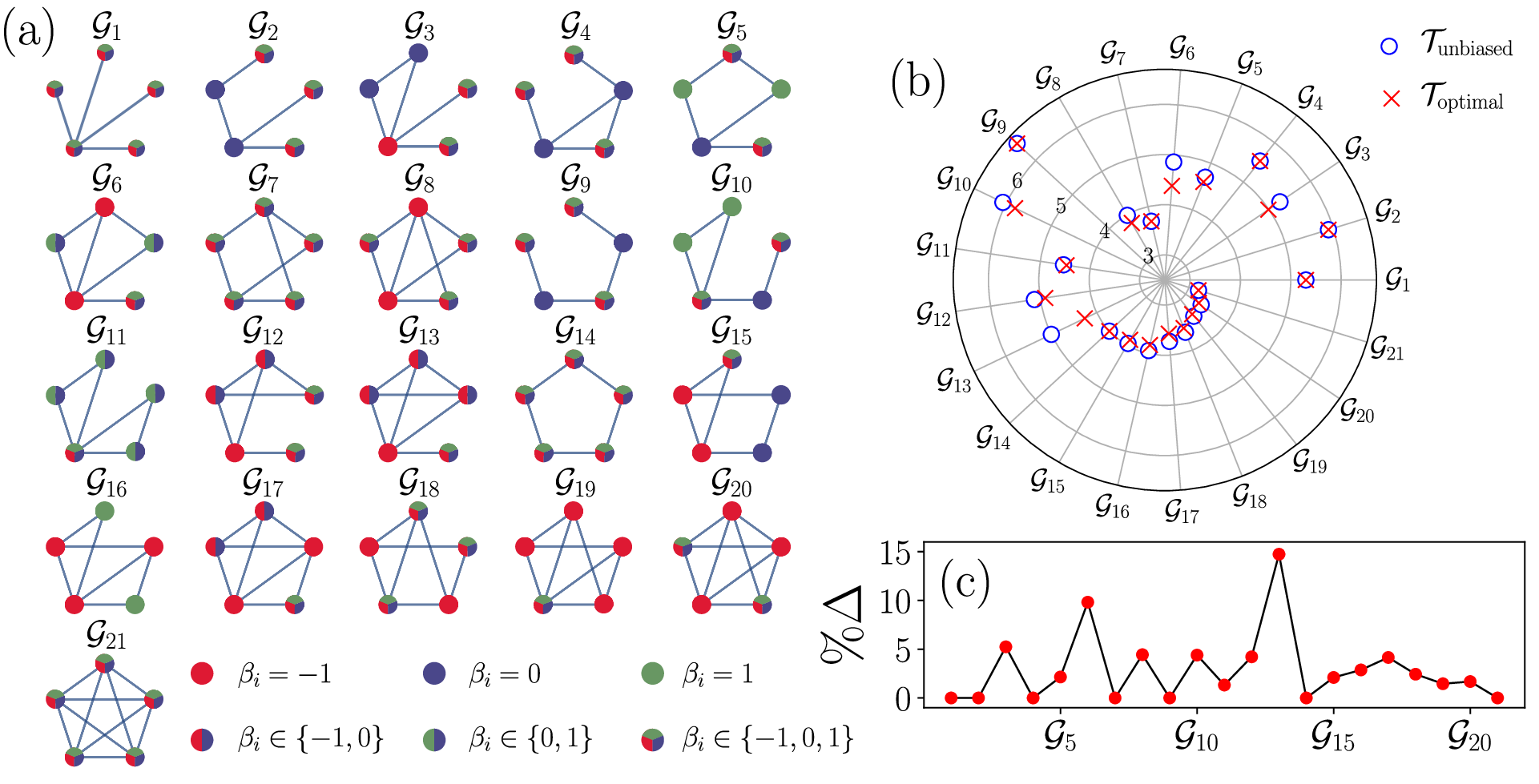}
		\caption{ \label{Fig_4} Random walks with local bias on graphs with $N=5$. (a) Multiplicity of $\vec{\beta}$ for the optimal configurations (with minimal  $\mathcal{T}=\mathcal{T}_{\mathrm{optimal}}$) for the
			$\mathcal{G}_1,\,\mathcal{G}_2,\ldots,\mathcal{G}_{21}$  non-isomorphic connected networks with $N=5$ nodes.  (b) Values of the optimal time $\mathcal{T}_{\mathrm{optimal}}$ and for the unbiased random walk $\mathcal{T}_{\mathrm{unbiased}}$  obtained for $\vec{\beta}=(0,0,0,0,0)$. (c) Difference between $\mathcal{T}_{\mathrm{optimal}}$ and $\mathcal{T}_{\mathrm{unbiased}}$ expressed as the percentage $\% \Delta$ defined in Eq. (\ref{Eq_Percent}).
		}
	\end{center}
\end{figure*}

In this section, we analyze random walks with local bias in graphs with $N=4$ and $N=5$ nodes. Due to their relative simplicity, we can explore the whole configuration space of $\vec{\beta}$ and find those that minimize the value of the global time  $\mathcal{T}$. For instance, the domain of $\vec{\beta}$ for a network with $N=4$ nodes has $3^4=81$ different configurations if $\beta_i\in\{-1,0,1\}$, whereas for $N=5$, $3^5=243$ configurations are possible. In these cases, the resulting random walk dynamics and their performance to explore the network can be evaluated efficiently using Eqs. (\ref{TauiSpect}) and (\ref{TauGlobal}). However, as the number of nodes in the graph increases, computing  $\mathcal{T}$ for each configuration becomes a difficult task. Our findings for small graphs are reported in Figs. \ref{Fig_3} and \ref{Fig_4}.
\\[2mm]
In Fig. \ref{Fig_3}, we present the analysis for all the non-isomorphic connected graphs with $N=4$. In Fig. \ref{Fig_3}(a), we show the values of the global time $\mathcal{T}$ (sorted in decreasing order) for each configuration of $\vec{\beta}$ for a particular network presented as inset. The global time $\mathcal{T}_{\mathrm{unbiased}}$ for the unbiased case is presented with a dashed line, three optimal configurations with  $\mathcal{T}_{\mathrm{optimal}}$ minimizing  $\mathcal{T}$ are shown with blue dots whereas the unbiased dynamics is presented with a black dot. Here, it is worth noticing that exploring the whole domain of $\vec{\beta}$, it is possible to find configurations more effective to reach all the nodes of the network than the unbiased random walk, these cases have values $\mathcal{T}<\mathcal{T}_{\mathrm{unbiased}}$, i.e., with values below the dashed line. Another important finding is that  optimal strategies for the local bias are not always unique. In Fig. \ref{Fig_3}(b), we illustrate the multiplicity of configurations with minimal $\mathcal{T}$  for all the non-isomorphic connected networks with $N=4$ (the 6 networks $\mathcal{G}_1,\,\mathcal{G}_2,\ldots,\mathcal{G}_6$ were obtained from \cite{ConnectedGraphs}), where we represent the graphs with different types of nodes according to the values $\vec{\beta}$ for the optimal configurations with $\mathcal{T}_{\mathrm{optimal}}$. For instance, in our example in inset (a) we have the graph $\mathcal{G}_3$. In this case, three optimal configurations with $\vec{\beta}=(\beta_1,\beta_2,\beta_3,\beta_4)$ are obtained with $\beta_1\in\{-1,0,1\}$,   $\beta_2=-1$,  $\beta_3=\beta_4=0$. 
\\[2mm]
In addition, in Fig. \ref{Fig_3}(b) we can see that there exists nodes unaltered by the bias. In these nodes (represented as pie charts) $\beta_i \in \{-1,0,1\}$. These particular nodes fulfill the following condition valid for connected undirected networks:
\begin{itemize}
\item[] If all the neighbors of the node $i$ have the same degree $k$, then
\begin{equation}\label{ConditionEqNeigh}
w_{i\rightarrow j}(\vec{\beta}\,)=\frac{A_{ij} k^{\beta_i}}{\sum_{l=1}^N A_{il} k^{\beta_i}}=\frac{A_{ij}}{k_i}.
\end{equation}
In this manner, the pass from $i$ to its neighbor $j$ is independent of $\beta_i$.
\end{itemize}
This condition is observed in two particular nodes in $\mathcal{G}_5$ in Fig. \ref{Fig_3}(b). In addition, if a node $i$ has only one neighbor, $k_i=1$, then
\begin{equation*}
w_{i\rightarrow j}(\vec{\beta}\,)=\frac{A_{ij} 1^{\beta_i}}{\sum_{l=1}^N A_{il} 1^{\beta_i}}=A_{ij}=1.
\end{equation*}
 This particular case is observed in graphs $\mathcal{G}_1$, $\mathcal{G}_2$, $\mathcal{G}_3$ in Fig. \ref{Fig_3}(b). On the other hand, in regular graphs all the nodes have the same degree $k$ fulfilling the condition (\ref{ConditionEqNeigh}). As a consequence the random walk is not affected by the vector $\vec{\beta}$. In Fig. \ref{Fig_3}(b), this is the case in $\mathcal{G}_4$ (ring with $k=2$) and $\mathcal{G}_6$ (fully connected graph with $k=3$). 
\\[2mm]
We complement the exploration of optimal configurations with the analysis of all the non-isomorphic connected networks with $N=5$ in Fig. \ref{Fig_4}, the graphs $\mathcal{G}_1,\,\mathcal{G}_2,\ldots,\mathcal{G}_{21}$ were obtained from \cite{ConnectedGraphs}. In Fig. \ref{Fig_4}(a), we present all the networks with different nodes representing the possible values of the bias $\beta_{i}$ in the optimal configurations. As a consequences of the condition (\ref{ConditionEqNeigh}) fulfilled for all the nodes, the dynamics on graphs $\mathcal{G}_1$, $\mathcal{G}_7$, $\mathcal{G}_{14}$ and $\mathcal{G}_{21}$ are not modified with the introduction of local bias. In other cases, optimal configurations occur with different biases and multiplicities represented with diverse types of nodes. In this respect, a pie chart denotes multiple values of $\beta_i$ for the bias minimizing the global time $\mathcal{T}$. Here, it is worth noticing that only $\mathcal{G}_{16}$ has a unique configuration for the optimal bias. 
\\[2mm]
On the other hand, in Fig. \ref{Fig_4}(b) we present a circular plot with the global times $\mathcal{T}_{\mathrm{optimal}}$ and $\mathcal{T}_{\mathrm{unbiased}}$. Finally, in Fig. \ref{Fig_4}(c), we present the relative difference between $\mathcal{T}_{\mathrm{optimal}}$ and $\mathcal{T}_{\mathrm{unbiased}}$
as a percentage given by
\begin{equation}\label{Eq_Percent}
\% \Delta\equiv \left(\frac{\mathcal{T}_{\mathrm{unbiased}}-\mathcal{T}_{\mathrm{optimal}}}{\mathcal{T}_{\mathrm{unbiased}}}\right)\times 100\% .
\end{equation}
We can see that in some cases, the improvement in the exploration of the network with the introduction of local bias is significant. For example, in $\mathcal{G}_6$ we have $\% \Delta=9.82\%$ whereas in $\mathcal{G}_{13}$, the optimal value is $\% \Delta=14.74\%$ better.
\\[2mm]
To summarize, we have proved that it is possible to find a better strategy than the unbiased case by adding a local bias to the random walker strategy, in some cases the optimal configurations are non unique. In the next section we expand our results to more complex and larger structures. 
\section{Optimal random walks}
\label{Optimal_Trans}
In small-size networks, we saw that it is possible to explore all the values of the vector $\vec{\beta}$ and identify configurations with the optimal local bias. However, if at each node $i$, $\beta_i$ can take $M$ values, the exploration of the complete domain $\vec{\beta}$ requires the evaluation of $M^N$ configurations. This exponential dependence with the size $N$ makes the exploration of optimal configurations that minimize $\mathcal{T}$  a difficult task for $N\gg 1$. In this section, we apply techniques for this minimization using random search algorithms developed to explore the configuration space efficiently to seek optimal ones. These type of algorithms have been applied on different branches of science like systems design \cite{kjellstrom1981stochastic}, operational research \cite{Eglese1990}, computational biology \cite{cedersund2016optimization}, spin glasses and quantum algorithms \cite{bapst2013quantum,PhysRevLett.104.207206}, machine learning \cite{GoodBengCour16}, among many others.
\\[2mm]
\begin{figure*}[t]
	\begin{center}
	    \includegraphics[width=0.95\linewidth]{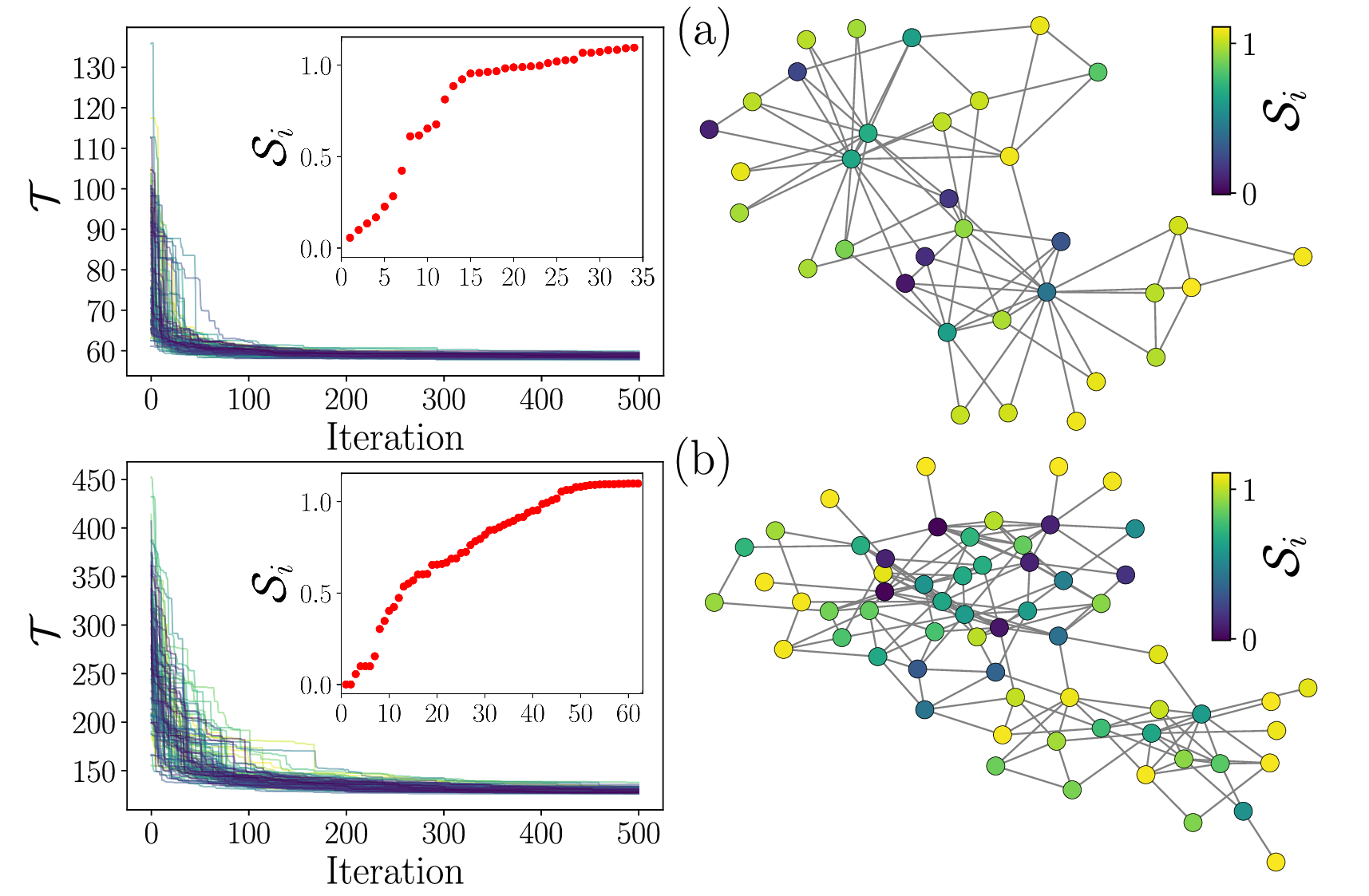} 
		\caption{\label{Fig_5} Monte Carlo simulations of the SA algorithm to optimize the transport with a local bias. (a) The Karate club network, (b) the Dolphins network.  In the left panels, we present the values $\mathcal{T}$ for  100 realizations with 500 iterations of the algorithm implemented with $\alpha=0.85$ and $T_0=1$. We show as an inset the results of the entropies found using  Eq. (\ref{Si_entropy}) for the optimal configurations, the results are sorted in increasing order. In the right panels, we depict each network analyzed with their nodes representing the entropy value encoded in the color bar. }
	\end{center}
\end{figure*}
In the following part, we implement the \textit{simulated annealing} (SA) algorithm, a Monte Carlo algorithm based on the physical process in which a solid is slowly cooled to reach a minimum energy configuration \cite{van1987simulated,berg1991multicanonical,hansmann1999generalized}. The basic idea of SA is to use random moves to reach minimum values of a cost function, such as the global time $\mathcal{T}$ as a function of $\vec{\beta}$, that physically plays the role of the system's energy with a transition probability given by the Boltzmann distribution law. At each move, we accept a new optimal configuration proposal $\vec{\beta}'$ with probability
\begin{equation}
    \rho=\min\left\{1,\exp\left[-\frac{\mathcal{T}(\vec{\beta}')-\mathcal{T}(\vec{\beta})}{T}  \right]\right\}.
\end{equation}
Here, $T$ is a parameter that emulates the temperature and it is reduced at each Monte Carlo iteration in such a way that the probability of accepting a new configuration that does not favor the minimization of the cost function decreases, the values $\mathcal{T}(\vec{\beta}')$ and $\mathcal{T}(\vec{\beta})$ are obtained numerically using Eqs. (\ref{TauiSpect}) and (\ref{TauGlobal}). The principal characteristic of SA is that it provides a form to escape from local minimums by occasionally accepting a configuration with a higher cost function. The algorithm has been widely used in a variety of problems such as clustering \cite{brown1992practical}, deep learning \cite{rere2015simulated}, quantum computing \cite{somma2008quantum}, chaotic systems \cite{mingjun2004application}, just to mention a few examples. 
\\[2mm]
The convergence of the algorithm mostly depends on the type of temperature decrease as well as the way the configurations are modified \cite{nourani1998comparison}. In our implementation of the algorithm, the temperature decreases logarithmically, that is
\begin{equation}
    T_\ell = \frac{T_0}{\ln(1+\ell)}, 
\end{equation}
\begin{table*}[t]
\begin{center}
	\begin{tabular}{c c c c c c c c c c c} 
		\hline 
		{\bf Network} & $N$ & $|\mathcal{E}|$ & $\bar{k}$ &$\mathcal{T}_{\mathrm{unbiased}}$  & $\left \langle\mathcal{T}_{\mathrm{optimal}}\right\rangle$ & $\sigma_{\mathcal{T}_{\mathrm{optimal}}}$  & $\bar{\mathcal{S}}$ & $\sigma_{S}$ &$\%\Delta$  \\[1mm]
		\hline 
		\hline
		Network 9 & 9 & 13 & 2.889 & 11.3864 & 10.522 & 0.018 & 0.668 & 0.456 & 7.59\\ 
		Network 10 & 10 & 21 & 4.2 & 13.5057 & 11.4096 & 0.02 & 0.561 & 0.5 & 15.52\\ 
		Karate club \cite{Zachary1977} & 34 & 78 & 4.588 & 63.916 & 58.554 & 0.488 & 0.794 & 0.334 & 8.39\\ 
		Dolphins \cite{bottlenose} & 62 & 159 & 5.129 & 155.231 & 129.479 & 2.273 & 0.756 & 0.325 & 16.59\\
		Erd\H{o}s-R\'enyi \cite{ErdosRenyi1959} & 50 & 69 & 2.76 & 143.055 & 131.97 & 1.827 & 0.875 & 0.253 & 7.75\\
		Watts-Strogatz \cite{WattsStrogatz1998} & 50 & 100 & 4.0 & 137.887 & 132.909 & 0.085 & 0.656 & 0.408 & 3.61\\
		Barab\'asi-Albert \cite{BarabasiAlbert1999} & 50 & 97 & 3.88 & 88.823 & 84.661 & 0.53 & 0.932 & 0.15 & 4.69\\
	Metro New York \cite{SybilWolfram2014,Derrible2012} & 77 & 109 & 2.831 & 280.02 & 260.369 & 4.894 & 0.685 & 0.374 & 7.02\\
		Metro Paris \cite{SybilWolfram2014,Derrible2012} & 78 & 125 & 3.205 & 239.196 & 227.816 & 2.479 & 0.714 & 0.404 & 4.76\\
		Metro London \cite{SybilWolfram2014,Derrible2012} & 83 & 121 & 2.916 & 329.138 & 311.495 & 4.709 & 0.721 & 0.386 & 5.36\\
		\hline
	\end{tabular}
\caption{\label{Table_1} Characterization of the optimal local bias obtained through the SA algorithm in different network topologies. For each network we present the number nodes $N$, the total number of edges $|\mathcal{E}|$, the average degree $\bar{k}$, the global time $\mathcal{T}_{\mathrm{unbiased}}$ and the values $\left \langle\mathcal{T}_{\mathrm{optimal}}\right\rangle$, $\sigma_{\mathcal{T}_{\mathrm{optimal}}}$,
$\bar{\mathcal{S}}$, $\sigma_{S}$, $\%\Delta$ characterizing the final results obtained with 100 Monte Carlo realizations of the SA algorithm implemented with 500 iterations, $\alpha=0.85$ and  $T_0=1$ (see details in the main text).} 
\end{center}
\end{table*}
\begin{figure}[!t] 
	\begin{center}
		\includegraphics*[width=0.48\textwidth]{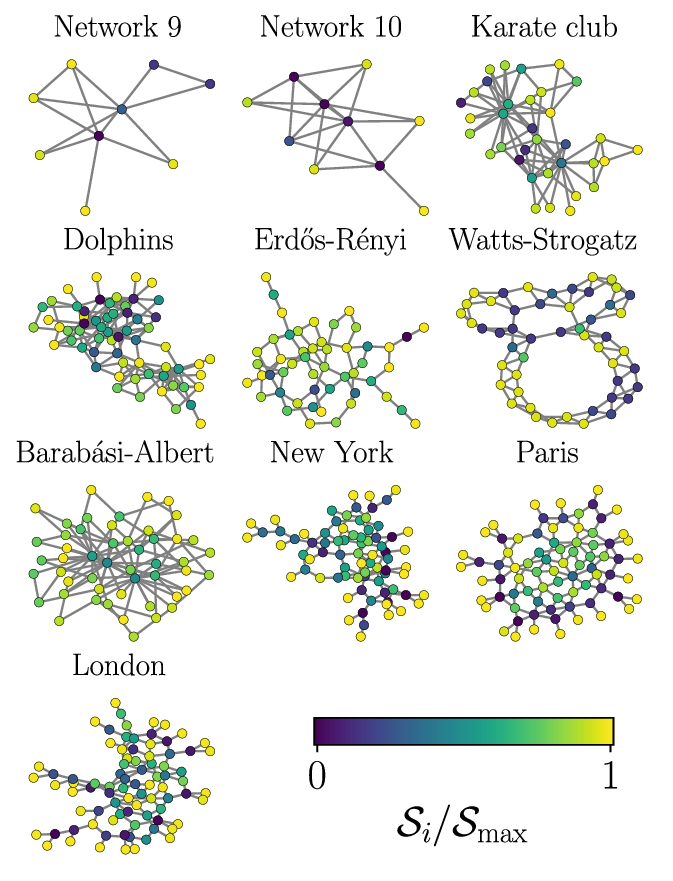}
	\end{center}
	\vspace{-5mm}
	\caption{\label{Fig_6} Networks analyzed in Table \ref{Table_1}. The nodes represent the value $\mathcal{S}_i/\mathcal{S}_{\mathrm{max}}$, with $\mathcal{S}_{\mathrm{max}}=\ln(3)$.  } 
\end{figure}
where $T_\ell$ is the temperature at iteration $\ell$ and $T_0$ is the initial temperature, we choose the initial configuration $\vec{\beta}$ with uniformly distributed random entries $\beta_i\in\{-1,0,1\}$. Moreover, the variation of a given configuration is performed according to a trigger parameter $\alpha$. Given an actual configuration $\vec{\beta}$, for each iteration we generate a random number $p\in (0,1)$ uniformly distributed; then, if  $p<\alpha$, only two entries are modified randomly and for $p \geq \alpha$, half of its entries change randomly. This second alternative allows the algorithm to continue exploring the whole configuration space and not to get stuck in a local minimum when the temperature is quite small. Note that, when $\alpha$ is close to one, the modifications of $\vec{\beta}$ lie around the current configuration except for rare occasions where the change is abrupt. In addition, based on the results in Sec. \ref{Sec_Small}, we only modify the entries of $\vec{\beta}$ assigned to nodes with degree $k>1$ or where their nearest neighbors have different degrees. In this manner, we center the implementation of the SA algorithm to produce modifications in the global time $\mathcal{T}$.
\\[2mm]
In Fig. \ref{Fig_5} we illustrate the implementation of the SA algorithm to find optimal configurations of the local bias $\vec{\beta}$ reducing the global time $\mathcal{T}$. We explore two networks: the \textit{Karate club network} \cite{Zachary1977} with $N=34$ presented in Fig. \ref{Fig_5}(a), and the \textit{Dolphins network} \cite{bottlenose}  with $N= 62 $ in Fig. \ref{Fig_5}(b). The results are obtained considering 100 Monte Carlo realizations with 500 iterations of the SA algorithm with $\alpha=0.85$ and $T_0=1$. Here, it is worth mentioning that in each realization, a different optimal value of $\mathcal{T}$ is obtained. This is due to the fact that SA is a heuristic method to get closer to a global minimum and obtain a similar result in each realization is unlikely if the configuration space is large where different minimums present multiplicities in the distribution of values $\beta_i$ as we illustrate in the analysis of small graphs in Figs. \ref{Fig_3}-\ref{Fig_4}.
\\[2mm]
Moreover, the uncertainty of the multiple configurations can be quantified associating to each node $i$ an entropy defined by
\begin{equation}\label{Si_entropy}
	\mathcal{S}_i = -\sum_{m\in\{-1,0,1\}}f^{(i)}_m\ln(f^{(i)}_m),
\end{equation}
where $f^{(i)}_m$ is the relative frequency of the values $m$ that the $i$-th entry of $\vec{\beta}$ adopts in the final configurations obtained with the SA algorithm. This compact representation of our results allows us to determine how well defined is the local bias in a particular node. On the left panels in Fig. \ref{Fig_5}, we present as an inset the values of  $\mathcal{S}_i$ found for the final configurations produced with the SA algorithm. On the right, we show the respective networks with the nodes colored according to their entropy. In particular, nodes with $\mathcal{S}_i=0$ are defined by a unique bias $\beta_i$ whereas $\mathcal{S}_i=\ln(3)$ indicates that $\beta_i$ can take with equal probability the values $\beta_i\in\{-1,0,1\}$.
\\[2mm]
The analysis made in Fig. \ref{Fig_5} can be implemented to other types of networks. In Table \ref{Table_1}, we expand our results to diverse graphs with different nature, topology, and size. In the first columns, we present a general description of the network features such as the size $N$, number of edges $|\mathcal{E}|=\frac{1}{2}\sum_{l,m}A_{lm}$ and average  degree $\bar{k}=\frac{1}{N}\sum_{i=1}^N k_i$. We also include $\mathcal{T}_{\mathrm{unbiased}}$
 for the unbiased case and the ensemble average $\left \langle\mathcal{T}_{\mathrm{optimal}}\right\rangle$
  for the optimal configurations found with the SA algorithm for  100 realizations with 500 iterations using $\alpha=0.85$ and $T_0=1$. The standard deviation of the results for all the realizations is denoted as $\sigma_{\mathcal{T}_{\mathrm{optimal}}}$. We also include the average of the entropy for all the nodes $\bar{\mathcal{S}}=\frac{1}{N}\sum_{i=1}^N \mathcal{S}_i$ and the respective standard deviation $\sigma_{S}$ for the values obtained in the nodes. In the last column, we present the relative difference $\% \Delta$ defined in Eq. (\ref{Eq_Percent}) to compare the unbiased case with the value $\left \langle\mathcal{T}_{\mathrm{optimal}}\right\rangle$. In Fig. \ref{Fig_6} we present the networks analyzed, the color of each node $i$ represents the normalized value $\mathcal{S}_i/\mathcal{S}_{\mathrm{max}}$, with $\mathcal{S}_{\mathrm{max}}=\ln(3)$.
\\[2mm]
The networks analyzed in Table \ref{Table_1} include two small size networks with $9$ and $10$ nodes, the Karate club and the Dolphins network analyzed in Fig. \ref{Fig_5}, three synthetic random networks generated with Erd\H{o}s-R\'enyi model \cite{ErdosRenyi1959}, the Watts-Strogatz algorithm \cite{WattsStrogatz1998} generated from a ring with nearest-neighbor and next-nearest-neighbor links and a rewiring probability of $p=0.05$ and a complex network of the Barab\'asi-Albert type, generated with the preferential attachment rule \cite{BarabasiAlbert1999}. 
\\[2mm]
In addition, in different urban transportation networks, the main function of the infrastructure is to communicate efficiently to all the nodes. In some cases, models with random walkers can help us to characterize the mobility in a particular transportation mode \cite{LoaizaMonsalvePlosOne2019,RiascosMateosSciRep2020,RiascosSandersPRE2021} or to understand the effect of damage \cite{erasohernandez2021}. Due to possible applications in the context of human mobility and urban transport, we analyze the metro networks in New York, Paris, and London. The networks were obtained from Refs. \cite{SybilWolfram2014,Derrible2012}, in this case, the nodes represent stations where users can change between lines or the end stations of a line (see details in Ref. \cite{Derrible2012}). 
\\[2mm]
The results for all the networks explored show that the SA algorithm allows obtaining local bias configurations with $\mathcal{T}<\mathcal{T}_{\mathrm{unbiased}}$, the values for the average over realizations $\langle\mathcal{T}_{\mathrm{optimal}}\rangle$ present small variations quantified with the standard deviation  $\sigma_{\mathcal{T}_{\mathrm{optimal}}}$. Although the final results for $\mathcal{T} $ are similar, the local bias configurations can be very varied, something that is observed with the entropy values $\bar{\mathcal{S}} $ averaged between all nodes, with standard deviations $\sigma_{S}$ that are high showing that there may exist nodes with well-defined local biases with their respective $\mathcal{S} _i\approx 0 $ or nodes where local bias does not affect dynamics with $\mathcal{S}_i\approx \ln(3)$. Regardless of this multiplicity, it is always possible to improve the walker without bias, something that is revealed in the relative differences measured with $\%\Delta$.
\\[2mm]
In this manner, the optimization of $\mathcal{T}$ obtained with the modification of a random walk strategy with local parameters is important because maintains the same adjacency matrix $\mathbf{A}$ without the introduction of new edges or rewiring of nodes, a fact that, in some systems may signify a huge increase of the operational costs; for example, the introduction of new lines in metro systems.  All the results in this section reveal that beneficial improvements in the bias can be obtained through the implementation of the SA algorithm while the resulting average entropy gives us an idea of the multiplicity that these strategies can have.

\section{Conclusions}
In this research, we introduce a random walk strategy with a local bias. In this case, transition probabilities between nodes are defined in terms of values $\beta_i$ in each node, a vector $\vec{\beta}=(\beta_1,\beta_2,\ldots,\beta_N)$ contains all the information of the biases. We explore the capacity of this random walker to visit the nodes of the network by using a global time $\mathcal{T}$ expressed in terms of eigenvalues and eigenvectors of the transition matrix defining the random walker.
\\[2mm]
Through the analysis of connected non-isomorphic graphs with $N=4, 5$ nodes, we explore all the possible configurations of local bias for $\beta_i\in \{-1,0,1\}$. The results reveal the existence of multiple configurations minimizing $\mathcal{T}$. In different graphs, optimal bias improves the capacity of the walker to explore the network in comparison with the unbiased dynamics (standard random walk). In other cases, the introduction of local bias does not alter the dynamics. We deduce rules to identify nodes where the local bias does not change the elements of the transition matrix.
\\[2mm]
As the identification of optimal configurations is a problem with multiple minimal, we implement a simulated annealing algorithm to explore configurations close to an optimal value of $\mathcal{T}$ for diverse types of real and synthetic networks with different sizes and topologies. This approach allows deducing configurations for which local biases improve the dynamics in comparison with the unbiased random walker. The multiplicity of the local bias in these configurations is quantified in terms of an entropy obtained from the statistical analysis of the results produced by multiple realizations of the algorithm.
\\[2mm]
The framework introduced in this research is general and provides a tool for the exploration of different random walk dynamics defined in terms of parameters taking values at each node. The understanding of stochastic processes with this characteristic and its optimization can be useful in different contexts, for example in routing processes as well as the planning of routes in urban transportation systems.
\section*{Acknowledgments}
CSHC acknowledges support from CONACYT M\'exico. APR acknowledges financial support from by PAPIIT-UNAM grant No. IN116220.

\onecolumngrid

\end{document}